\documentclass[reprint,aps,ams,amsmath,prx,twocolumn,superscriptaddress,longbibliography]{revtex4-1}
\usepackage{graphics}
\usepackage{graphicx}
\usepackage{epsfig}
\usepackage{amsmath}
\usepackage{amssymb}
\usepackage{amsfonts}
\usepackage{bm}
\usepackage{color}
\usepackage{bbm}
\usepackage{hyperref}
\usepackage[normalem]{ulem}
\usepackage{comment}
\usepackage{soul}
\DeclareMathOperator{\ReP}{Re}
\DeclareMathOperator{\ImP}{Im}

\begin{document}

\title{Fluctuation-driven thermal transport in graphene double-layers at charge neutrality}

\author{Alex Levchenko}
\affiliation{Department of Physics, University of Wisconsin-Madison, Madison, Wisconsin 53706, USA}

\author{Songci Li}
\affiliation{Department of Physics, University of Wisconsin-Madison, Madison, Wisconsin 53706, USA}

\author{A. V. Andreev}
\affiliation{Department of Physics, University of Washington, Seattle, Washington 98195, USA}

\date{September 10, 2022}

\begin{abstract}
We develop a theory of fluctuation-driven phenomena in thermal transport in graphene double-layers. We work in the regime of electron hydrodynamics and focus on the double-charge neutrality point. Although at the neutrality point charge transport is decoupled from the hydrodynamic flow, thermal fluctuations of electron density cause both drag and heat transfer between the layers. The thermal transport in the bilayer system is governed by these two phenomena. We express the drag friction coefficient and the interlayer thermal conductivity in terms of the interlayer distance and the intrinsic conductivity of the electron liquid. We then obtain the thermal conductance matrix and determine the spatial dependence of the hydrodynamic velocity and temperature in the system. For shorter system the thermal drag resistance is determined by drag. In longer systems the situation of perfect thermal drag is realized, in which the hydrodynamic velocities in both layers become equal in the interior of the systems. Estimates are given for the monolayer and bilayer graphene devices. The predictions of our theory can be tested by the high-resolution thermal imaging and Johnson-Nyquist nonlocal noise thermometry.
\end{abstract}

\maketitle

\section{Introduction}

The electronic-double-layers (EDL) -- spatially separated conducting electron systems -- provide a versatile platform to study quantum effects of electronic correlations. In these systems, the coupling strength between the layers can be effectively controlled by changing the interlayer separation and/or carrier concentration, and the resulting state of the system can be measured in a broad range of temperatures and magnetic fields. As a consequence of this tunability, the interlayer electron-electron interactions may lead to a number of interesting phenomena that include, for example, interlayer excitonic superfluidity \cite{Eisenstein,MacDonald}, even-denominator fractional quantum Hall states \cite{Tsui,Eisenstein-He,Luhman}, and nonlocal frictional transport effects of Coulomb drag \cite{Rojo,RMP-Drag}.

In recent years, the great interest in EDL has been motivated by the advent of graphene-boron-nitride heterostructures \cite{Ponomarenko,Britnell}, which include both monolayer graphene (MLG) \cite{Tutuc,Kim-Tutuc,Geim-NP12,Geim-PRL13} and bilayer graphene (BLG) \cite{Morpurgo,Dean,Hone} double-layers. These devices offer a number of advantages compared with quantum wells of GaAlAs-based two-dimensional electron systems. It takes only a few atomic layers of hexagonal boron nitride (hBN) to confine carriers in graphene within a single atomic plane and isolate graphene electrically. This allows to create double-layers with unprecedentedly small interlayer separation and thus strong Coulomb coupling, since hBN has relatively small dielectric constant $(\varepsilon\approx4)$. The charge carriers within each layer can be independently and continuously tuned in a broad range of densities from the charge-neutral state to high density of either electrons or holes. Therefore, the Coulomb drag in graphene EDL heterostructures can be studied not only in the temperature regime of degenerate electron gas $T<E_F$, where $E_F$ is the Fermi energy, but also in the regime of correlated fluid at higher temperatures, where electronic system attains hydrodynamic limit \cite{NGMS,Lucas-Fong,ALJS}. 

In this paper we study hydrodynamic thermal transport in a double-layer system at the double charge neutrality point, where the charge density in each layer is zero. The most salient feature of electron hydrodynamics at charge neutrality is the decoupling of the hydrodynamic flow from the charge current. This decoupling holds only on average, while fluctuations of the electron density produce mixing of charge current and the hydrodynamic flow. 
Thermal charge fluctuations transfer both energy and momentum between the layers, causing both thermal drag~\cite{Andreev1972,Fazio,Moore,Ben-Abdallah2019,Sukhorukov} and interlayer thermal conductivity~\cite{Polder-VanHove,Pendry,Mahan,Kamenev-NFHT,Basko-NFHT,Principi-Polini,Ying-Kamenev}, for the latter see also reviews Refs. \cite{Volokitin-Persson,Ben-Abdallah} devoted to the near-field heat transfer (NFHT). Both of these fluctuation-driven phenomena modify the hydrodynamic flow in both layers and thereby determine the thermal transport properties of the double-layer system. We treat the thermal fluctuations by introducing the Langevin sources into the hydrodynamic equations. This is done by a straightforward extension of Landau and Lifshitz's theory~\cite{LL-JETP57} to non-Galilean-invariant liquids. The particular advantage of hydrodynamic approach is that it enables considerations beyond the perturbation theory in interaction. We show that, at charge neutrality, the thermal drag resistance and NFHT conductance can be expressed in terms of intrinsic conductivity of the pristine fluid and its thermodynamic entropy density. We also work out a generic four-terminal setup and calculate spatial distribution profiles for the electronic temperature in the EDL, which can be mapped out experimentally via high-resolution thermal imaging probes \cite{Zeldov} and Johnson-Nyquist nonlocal noise thermometry \cite{Waissman}. 

The  paper is organized as follows. In Sec. \ref{sec:hydro} we develop a Langevin treatment of hydrodynamic fluctuations applicable to non-Galilean invariant electron liquids.  In Sec. \ref{sec:drag} we apply this approach to evaluate the thermal drag coefficient and interlayer thermal conductivity. Using these results we develop a macroscopic theory of thermal transport in electron bilayers, and provide experimentally relevant estimates for MLG and BLG devices. In Sec. \ref{sec:summary} we summarize the main findings of this work and outline perspectives for further extensions. In Appendix \ref{sec:appendix} we outline an alternative approach to the treatment of thermal charge fluctuations, which represents an extension of the Langevin theory of van der Waals forces developed by E. M. Lifshitz~\cite{lifshitz1955theory,lifshitz1992theory} to nonequilibrium quantities. This approach does not rely on the hydrodynamic approximation, and is applicable to electron layers of arbitrary thickness.  

\section{Hydrodynamic fluctuations}\label{sec:hydro}

In this section we formulate the transport theory of hydrodynamic fluctuations in electron liquids. For that purpose, we follow the classic work Ref. \cite{LL-JETP57} making the necessary generalizations to account for the presence of Coulomb interaction and broken Galilean invariance of electron liquids in solids. 

The hydrodynamic equations express conservation of the density of particle number, energy, and momentum of the electron liquid. Since the energy density depends on the entropy density of the liquid the energy conservation equation in hydrodynamics is traditionally replaced by an equivalent evolution equation for the entropy density \cite{LL-V6}. Since entropy production is quadratic in deviations from equilibrium, it may be neglected for the purpose of studying linear transport. In this approximation the entropy evolution equation expresses conservation of entropy. Accordingly, the time derivatives of the number density $n$, entropy density $s$, and momentum density $p_i$ may be expressed in terms of divergences of the corresponding conserved fluxes $\bm{j}_n$, $\bm{j}_s$, and $\Pi_{ij}$. To keep subsequent expressions more compact, it is convenient to combine the thermodynamic densities as well as particle and entropy fluxes into two-component column vectors
\begin{equation}\label{eq:x-J}
\vec{x}=\left(\begin{array}{c}n \\ s\end{array}\right),\quad \vec{\bm{J}}=\left(\begin{array}{c}\bm{j}_n \\ \bm{j}_s\end{array}\right).
\end{equation}
Here and in what follows the column vector quantities are denoted by arrows above them, and we use
bold face letters to denote the usual spatial vectors. In the notations of Eq.~\eqref{eq:x-J}, conservation of particle number  and entropy  are expressed by the continuity equation 
\begin{equation}\label{eq:dt-x}
\partial_t\vec{x}=-\bm{\nabla}\cdot\vec{\bm{J}},
\end{equation}
while the evolution equation for the momentum density has the form of Newton's second law,
\begin{equation}\label{eq:dt-p}
\partial_t\bm{p}=-\bm{\nabla}\cdot\hat{\Pi}-en\bm{\nabla}\phi.
\end{equation}
The electric potential $\phi$ here is related to the electron density by the Poisson equation. Its presence reflects the flow of momentum of the electron fluid due
to long-range Coulomb interactions between electrons, whereas the tensor $\hat{\Pi}\equiv\Pi_{ij}$ denotes the local part of the momentum flux.

An essential ingredient of the hydrodynamic approach is the assumption of local thermal equilibrium of the electron liquid. Accordingly, the state of the liquid is characterized by the local equilibrium parameters: temperature $T$, chemical potential $\mu$, and the hydrodynamic velocity $\bm{v}$, whose values are determined by the local densities of conserved quantities.  In the hydrodynamic approximation the fluxes of conserved quantities are expanded to first order in the gradients of equilibrium parameters.

Hydrodynamic fluctuations are described using the Langevin approach by adding fluctuation terms to the hydrodynamic constitutive relations for the conserved currents~\cite{LL-JETP57}. For liquids that do not possess Galilean invariance~\footnote{In Galilean-invariant liquids, the particle current density is uniquely determined by the local hydrodynamic velocity $\bm{v}$, which in turn is defined by the momentum density. In the absence of Galilean invariance, additional and dissipative contributions of the particle current arises.}, the expressions for the currents in Eq.~\eqref{eq:dt-x} take the form 
\begin{equation}\label{eq:J}
\vec{\bm{J}}=\vec{x}\bm{v}-\hat{\Upsilon}\vec{\bm{X}}+\vec{\bm{J}}_L.
\end{equation}
The first term on the right-hand-side of Eq.~\eqref{eq:J} describes the convective transport of charge and entropy by the hydrodynamic flow with the hydrodynamic velocity $\bm{v}$. The second term describes dissipative transport of charge and heat relative to the fluid in response to driving forces $\vec{\bm{X}}$, which are thermodynamically conjugate to the corresponding densities $\vec{x}$ \cite{LL-V5}. Specifically, the column-vector $\vec{\bm{X}}^{\mathbb{T}}=(-e\bm{\mathcal{E}},\bm{\nabla}T)$ consists of the electromotive force, $e\bm{\mathcal{E}}=-\bm{\nabla}(\mu+e\phi)$, and the local temperature gradients, $\bm{\nabla}T$, generated in a fluid (above we used symbol $\mathbb{T}$ to denote vector transposition). The matrix of kinetic coefficients $\hat{\Upsilon}$ characterizes the dissipative properties of the electron liquid. It is given by
\begin{equation}\label{eq:Upsilon}
\hat{\Upsilon}=\left(\begin{array}{cc}\sigma/e^2 & \gamma/T \\ \gamma/T & \kappa/T\end{array}\right),
\end{equation}
where $\kappa$ is the thermal conductivity, $\sigma$ is the intrinsic conductivity, and $\gamma$ is the thermoelectric coefficient of
the electron liquid. The assumption of broken Galilean invariance is manifested by nonvanishing $\sigma$ and $\gamma$. The third term on the right-hand-side  of Eq.~\eqref{eq:J} captures the Langevin currents $\vec{\bm{J}}_L$ that describe thermally driven spatial and temporal fluctuations whose variances are related to dissipative transport coefficients by the fluctuation-dissipation theorem \cite{LL-V9,Kogan}
\begin{equation}\label{eq:J-J}
\left\langle\vec{\bm{J}}_L(\bm{r},t)\otimes\vec{\bm{J}}^{\mathbb{T}}_L(\bm{r}',t')\right\rangle=2T\hat{\Upsilon}\delta(\bm{r}-\bm{r}')\delta(t-t'). 
\end{equation}
The notation $\vec{a}\otimes\vec{b}^{\mathbb{T}}$ is used to denote the direct product of two vectors.

The momentum flux tensor of the electron liquid in Eq.~\eqref{eq:dt-p},
\begin{equation}
\Pi_{ij}=P\delta_{ij}-\Sigma_{ij},
\end{equation}
comprises the local hydrodynamic pressure $P$ and viscous stress tensor
\begin{equation}
\Sigma_{ij}=\eta(\partial_iv_j+\partial_jv_i)+(\zeta-\eta)\delta_{ij}\partial_kv_k+\varsigma_{ik},
\end{equation}
where $\eta$ and $\zeta$ are, respectively, the shear and bulk viscosities. The form of $\Sigma_{ij}$ was tailored to the two spatial dimensions. The last term in the definition of $\Sigma_{ij}$ denotes stochastic Langevin viscous stresses, whose correlation function is given by \cite{LL-JETP57}
\begin{align}\label{eq:zeta-zeta}
\left\langle\varsigma_{ik}(\bm{r},t)\varsigma_{lm}(\bm{r}',t')\right\rangle & = [\eta(\delta_{il}\delta_{km}+\delta_{im}\delta_{kl})+(\zeta-\eta)\delta_{ik}\delta_{lm}]  \nonumber \\ 
&  \times 2T  \delta(\bm{r}-\bm{r}')\delta(t-t') .
\end{align}

The Langevin scheme outlined above provides a description of fluctuation-driven phenomena in the hydrodynamic regime. It assumes local, but not global thermal equilibrium, and therefore remains applicable in the presence of a hydrodynamic flow. For a given device geometry these equations need to be supplemented by the appropriate boundary conditions. As we show below, this enables evaluation of both fluctuation correlation functions, and thermoelectric transport coefficients, which are affected by thermal fluctuations. 

\begin{figure}[t!]
\includegraphics[width=\linewidth]{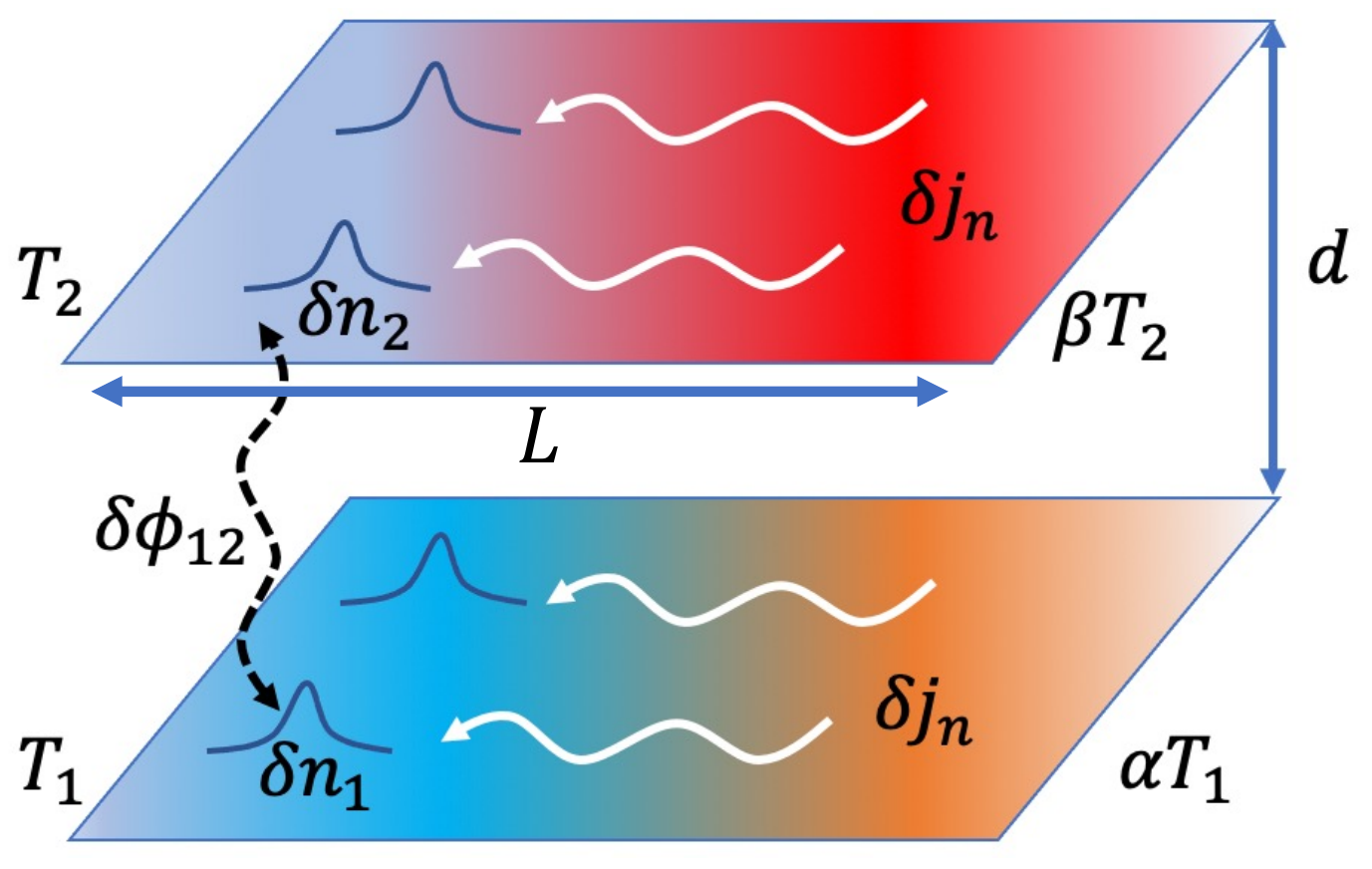}
\caption{A schematic representation of an electronic double-layer in the regime of thermal Coulomb drag with generic four-terminal biasing, which creates inhomogeneous temperature profiles $T_{1,2}(\bm{r})$. Thermal fluctuations of electron density  $\delta n_{1,2}(\bm{r},t)$ are advected by the hydrodynamic flow of heat. This results in momentum transfer between the layers, causing thermal drag. The drag force must be balanced by thermally-induced pressure gradient. This, in turns causes near-field transfer of thermal energy between the layers.   The relevant system dimensions include interlayer spacing $d$, layer length $L$, and thermal equilibration healing length $l$, whose value is detemined by the competition between the thermal drag and interlayer heat transfer [see Eq. \eqref{eq:l}].}  
\label{fig:EDL}
\end{figure}

\section{Thermal drag effect}\label{sec:drag}

In this section we use the theory of hydrodynamic fluctuations to study nonlocal thermal drag in mesoscopic EDL devices, and evaluate
thermal transport coefficients of these systems. A similar approach was recenly applied to the problem of Coulomb drag resistance~\cite{Apostolov-PRB14,Chen-PRB15,Patel,Apostolov-PRB19}. 

We consider a system formed by two  two-dimensional electron layers separated by a distance $d$, see Fig. \ref{fig:EDL}. For simplicity, we assume that both conducting layers are identical. To ensure applicability of hydrodynamic description, we must requite the interlayer separation to exceed the equilibration length of the electron liquid. For the electron liquid in graphene at charge neutrality, the latter is  on the order of
the thermal de Broglie  length, $\lambda_T$.
We focus on the most interesting regime of double charge neutrality, where each layer is charge neutral on average. In this case the hydrodynamic flow is decoupled from charge transport and corresponds to purely thermal flux. However, this decoupling holds only on average, whereas thermal fluctuations of the electron density violate it. 
In the presence of hydrodynamic flow velocity $\bm{v}$ in the active layer, the fluctuations of electron density with the wavevector $\bm{q}$ are advected by the flow. As a result, their spreading becomes anisotropic with respect to the direction of $\bm{v}$. The interlayer coupling of density fluctuations causes transfer of both energy and momentum between the layers, resulting in thermal drag. Below, we evaluate the rate of energy and momentum between the layers by neglecting electron-phonon coupling and accounting for the interlayer coupling of density fluctuations caused by the Coulomb interaction. 

\subsection{Drag force from the density fluctuations} 

We begin with the consideration of drag in a pristine systems at double charge neutrality. The key point to realize is that at charge neutrality the thermal fluctuations of charge density $\delta n$, entropy $\delta s$, and hydrodynamic velocity $\delta\bm{v}$ are independent to linear order. Moreover, the longitudinal fluctuations of velocity are coupled to pressure and propagate in the form of sound waves, while the transverse fluctuations of velocity spread via vorticity diffusion. To determine the drag force $\bm{\mathcal{F}}$, we thus focus on the density fluctuations and linearize Eq.~\eqref{eq:dt-x}. For the active layer, labeled by the subscript $1$,  we obtain 
\begin{equation}\label{eq:delta-n}
\partial_t\delta n_1+\bm{v}\cdot\bm{\nabla}\delta n_1-\frac{\sigma}{e^2}\bm{\nabla}^2(e\delta\phi_{12})+\bm{\nabla}\cdot\delta\bm{j}_{n1}=0,
\end{equation}
where $\delta\bm{j}_{n}$ is the part of the Langevin particle current, whose variance is proportional to the intrinsic conductivity in Eq. \eqref{eq:J-J}, and $\delta\phi_{12}$ is the Coulomb potential that includes density fluctuations in the passive layer.  In the passive layer, which is labeled by the subscript 2, the density fluctuations $\delta n_2$ are given by the same equation as Eq.~\eqref{eq:delta-n}, except that $\bm{v}=0$ and one needs to interchange the layer index $1\leftrightarrow2$. Passing to the Fourier transform, $\delta n\propto \exp(-i\omega t+i\bm{qr})$, and symmetrizing fluctuating density modes, $\delta n_{\pm}=\delta n_1\pm\delta n_2$, and corresponding fluctuating currents, $\delta\bm{j}_{n\pm}=\delta\bm{j}_{n1}\pm\delta\bm{j}_{n2}$, we obtain 
\begin{equation}\label{eq:delta-n}
\Gamma_{\pm}\delta n_\pm=-i(\bm{q}\cdot\delta\bm{j}_{n\pm})-\frac{i}{2}(\bm{q}\cdot\bm{v})(\delta n_++\delta n_-),
\end{equation}
where we introduced the susceptibility function 
\begin{equation}
\Gamma_\pm(q,\omega)=-i\omega+(2\pi\sigma q/\varepsilon)(1\pm e^{-qd}).
\end{equation}
To arrive at  Eq.~\eqref{eq:delta-n}  we used 
\begin{equation}
\delta\phi_{12}=\frac{2\pi e}{\varepsilon q}(\delta n_1+\delta n_2e^{-qd}), 
\end{equation}
where $\varepsilon$ is the dielectric constant of the medium surrounding the electron layers. 

In equilibrium, $\bm{v}=0$, fluctuating currents $\delta\bm{j}_{n\pm}$ render density fluctuations $\delta n_\pm$. These density modes are overdamped due to the finite intrinsic conductivity. Fluctuations relax exponentially in time, $\delta n_\pm\propto e^{-t\gamma_\pm}$, with the corresponding characteristic Maxwell rate \cite{Falko}
\begin{equation}\label{eq:tau}
\gamma_\pm(q)=(2\pi\sigma q/\varepsilon)(1\pm e^{-qd}).  
\end{equation}
This behavior should be contrasted with the hydrodynamic fluctuation dynamics at high density.  In that limit, density fluctuations, and thus drag effect, are dominated by the symmetric and antisymmetric plasmon modes, whose attenuation is governed by the fluid viscosity \cite{Apostolov-PRB14}.  

In the presence of the hydrodynamic flow in the active layer the correlation functions of the Langevin currents in Eqs.~\eqref{eq:J-J} and \eqref{eq:zeta-zeta} do not change, but the propagation of fluctuations is affected by the flow velocity $\bm{v}$. We account  for this
change to linear order in $\bm{v}$ by  splitting the density fluctuations $\delta n=\delta n^{(0)}+\delta n^{(1)}$ into the equilibrium part, $\delta n^{(0)}$, and the linear in $\bm{v}$ correction, $\delta n^{(1)}$. We thus find from Eq.~\eqref{eq:delta-n}
\begin{equation}\label{eq:delta-n-sol}
\delta n^{(0)}_{\pm}=\frac{\bm{q}\cdot\delta\bm{j}_{n\pm}}{ i \Gamma_\pm},\quad 
\delta n^{(1)}_{\pm}=\frac{\bm{q}\cdot\bm{v}}{i \Gamma_\pm}(\delta n^{(0)}_++\delta n^{(0)}_{-}).
\end{equation} 
These expressions, combined with Eq.~\eqref{eq:J-J}, enable evaluation of the drag force. The latter represents the average force exerted by density fluctuations in the active layer onto the passive layer, and is expressed by the following correlation function:
\begin{equation}\label{eq:F}
\bm{\mathcal{F}}=\int\frac{d^2qd\omega}{(2\pi)^3}(-i\bm{q})\left(\frac{2\pi e^2}{\varepsilon q}\right)e^{-qd}\mathcal{D}_F(\bm{q},\omega),
\end{equation}
where we introduced a shorthand notation  
\begin{equation}
\mathcal{D}_F(\bm{q},\omega)=\left\langle \delta n_1(\bm{q},\omega)\delta n_2(-\bm{q},-\omega)\right\rangle.
\end{equation}
Evaluating $\mathcal{D}_F$ to the linear order in $\bm{v}$, we observe that density averages of the same parity $\left\langle \delta n_\pm\delta n_\pm\right\rangle$ contain an overall factor $\propto (\Gamma_\pm-\Gamma^*_\pm)=-2i\omega$. Therefore, these terms are odd in frequency and consequently drop out from the force in Eq. \eqref{eq:F}. In contrast, contributions from the density averages of the opposite parity $\left\langle \delta n_\pm\delta n_\mp\right\rangle$ contain a factor of $\Gamma_\pm+\Gamma^*_\pm$, and thus are even in frequency. Collecting all the terms together, and using the variance of Langevin currents from Eq.~\eqref{eq:J-J},
\begin{equation}\label{eq:j-j}
\left\langle(\bm{q}\cdot\delta\bm{j}_{n\pm})(\bm{q}\cdot\delta\bm{j}_{n\pm})\right\rangle=4Tq^2\sigma/e^2,
\end{equation}   
we arrive at the following expression:
\begin{equation}\label{eq:D}
\mathcal{D}_F(\bm{q},\omega)=i(\bm{q}\cdot\bm{v}) \left(\frac{Tq^2\sigma}{e^2}\right)
\frac{\ReP (\Gamma_+ -\Gamma_-)}{|\Gamma_+|^2|\Gamma_-|^2}.
\end{equation}
We substitute this expression into Eq.~\eqref{eq:F} and 
evaluate the frequency integral first in the resulting expression, 
\[
\int^{+\infty}_{-\infty}\frac{d\omega}{|\Gamma_+|^2|\Gamma_-|^2}=\frac{\pi}{\gamma_+\gamma_-(\gamma_++\gamma_-)}.
\]
The integration over the directions of $\bm{q}$ in the remaining expression can be easily done and yields the drag force in the form 
\begin{equation}\label{eq:Fkth}
\bm{\mathcal{F}}=k_{\text{th}}\bm{v},
\end{equation}
were the coefficient of drag friction  $k_{\text{th}}$ is given by
\begin{equation} 
\label{eq:k_th_expression}
k_{\text{th}}=\frac{T\sigma}{4e^2}\int\frac{q^4d^2q}{4\pi^2}\left(\frac{2\pi e^2}{\varepsilon q}\right)e^{-qd}\frac{(\gamma_+-\gamma_-)}{\gamma_+\gamma_-(\gamma_++\gamma_-)}.
\end{equation} 
With the aid of  Eq.~\eqref{eq:tau}, the remaining momentum integral can be evaluated by introducing a dimensionless variable $x =2qd$. It brings the numerical factor $\int^{\infty}_{0}x^2e^{-x}dx/\sinh(x)=\zeta(3)/2$, where $\zeta(z)$ is the Riemann's zeta function. The resulting drag friction coefficient, 
\begin{equation}\label{eq:kth-result}
k_{\text{th}}=\frac{\zeta(3)}{64\pi^2} \left(\frac{\varepsilon T}{\sigma d^3}\right) ,
\end{equation} 
is inversely proportional to the third power of the interlayer separation $d$. Notice the unusual dependence on the dielectric constant in $k_{\text{th}}$, which originates from two competing effects. On the one hand, the dielectric environment weakens Coulomb coupling; on the other hand, it prolongs the relaxation time of charge fluctuation that contributes to the density correlation function and enhances the drag force.   

Note that, in the ideal liquid approximation where all the dissipative coefficients of the electron liquid are sent to zero, the  drag coefficient in Eq.~\eqref{eq:kth-result} diverges.  This may seem counterintuitive, as one might naively expect that, in the ideal liquid approximation, the drag coefficient and resistivity must vanish. This expectation is however incorrect because it is based on the implicit assumption that the charge density in the two layers is uniform. However, this is true only on average, whereas the equal-time density fluctuations, which depend only on thermodynamic quantities (compressibility) remain nonzero in the ideal liquid limit. Furthermore, in the ideal liquid approximation all relaxation processes are switched off. Therefore these equal time fluctuations remain ``frozen" into the liquid.  This results in perfect linear response drag. Indeed, equal-time density fluctuations are correlated between the layers. Therefore dragging one liquid  past the other in the presence of frozen-in density variations requires overcoming a finite energy barrier, which arises from interlayer correlations of equal-time density fluctuations.  A similar situation arises in the hydrodynamic treatment of resistivity in systems subjected to long disorder~\cite{AKS}. In that case the resistivity is inversely proportional to the intrinsic thermal conductivity, and also diverges in the ideal liquid limit. In that case the liquid is not isentropic in equilibrium because the entropy per particle depends on the local disorder potential. Because of that, initiation of the flow requires overcoming a finite energy barrier (stagnation enthalpy~\cite{Batchelor}), and linear-response flow is impossible. Note that both of these states (perfect drag, and stagnant liquid) are dissipationless. In the case of drag we have perfect locking of the layers coupled by elastic forces. In the case of stagnant liquid the liquid is locked to disorder at small drives.

It may be instructive to contrast this result with the earlier calculation of $k_{\text{th}}$ carried out for the case of Galilean-invariant liquids  \cite{Apostolov-PRB14}. In that case the hydrodynamic density fluctuations consist of plasmons driven by fluctuating longitudinal stresses, diffusive modes caused by temperature fluctuations, and thermal expansion of the electron liquid. It turns out that the plasmons give the dominant contribution to drag in the parameter $k_Fd\gg1$, where $k_F$ is the Fermi wave number. The resulting drag friction coefficient can be expressed as follows:
\begin{equation}\label{eq:kth-plasmons}
k_{\text{th}}\simeq \frac{\varepsilon T(\eta+\zeta)}{e^2n^2d^5}\ln^4\left[\frac{(nd)^3}{a_B(\eta+\zeta)}\right],   
\end{equation}
where $a_B$ is the effective Bohr radius in the material. For the Fermi liquid, $\eta\sim n(E_F/T)^2$, while bulk viscosity vanishes for a simple parabolic band or in general remains smaller than $\eta$ \cite{LL-V10}, therefore Eq. \eqref{eq:kth-plasmons} can be rewritten as 
\begin{equation}\label{eq:kth-Fermi}
k_{\text{th}}\sim \left(\frac{v_F\varepsilon}{e^2}\right)
\left(\frac{1}{k_Fd^5}\right)\frac{E_F}{T}\ln^4\left[\frac{E_F}{T}\frac{\sqrt[4]{a_B/d}}{\sqrt{k_Fd}}\right].    
\end{equation}
The applicability of this result is restricted to the temperature interval $E_F/\sqrt[4]{k_Fd}<T<E_F$ provided $k_Fd\gg1$. 

The comparison between Eq.~\eqref{eq:kth-result}, and Eqs.~\eqref{eq:kth-plasmons} and \eqref{eq:kth-Fermi} shows that the Coulomb drag in Dirac liquids is much stronger than that in Galilean-invariant liquids, and exhibits a slower ($1/d^3$ rather than $1/d^5$) falloff with interlayer distance. We expect that at high electron density (in the quantum degenerate regime), the Coulomb drag in Dirac liquids should approach the Galilean-invariant result. The study of the crossover between the regimes of high and low electron density is beyond the scope of the present work. 

\subsection{Thermal drag in short systems}

The drag force, characterized by the friction coefficient $k_{\text{th}}$ in Eq.~\eqref{eq:kth-result}, affects the hydrodynamic flow of the electron liquid. Since at charge neutrality hydrodynamic flow transports heat rather than charge, the drag force produces thermal drag resistivity. At charge neutrality the drag force must be balanced by the thermally induced pressure gradients arising within the layers, 
\begin{align}
    \label{eq:Tgrad_short}
    \bm{\mathcal{F}}=s\bm{\nabla}T_2 = -s\bm{\nabla}T_1.
\end{align}
Because of the opposite  sign of the temperature gradients in the active and passive layer, a position-dependent temperature difference between the two layers arises. In this situation, thermal fluctuations of the electron density produce not only momentum exchange but also energy exchange between the layers. 

In sufficiently short systems, the accumulated interlayer temperature difference caused by drag is small, and may be neglected. In this approximation thermal energy transfer between the layers is absent, in complete analogy with charge drag in which interlayer charge transfer is absent. The drag thermal resistivity, $\rho_{\text{th}}$, may be defined as the ratio of the temperature gradient induced in the passive layer to the conserved heat current, $\bm{j}_{q1}=T\bm{j}_{s1}$, in the active layer. Since the latter is given by  $\bm{j}_{q1}=Ts\bm{v}_1 - \kappa \bm{\nabla}T_1$, we get
\begin{equation}\label{eq:rth-intermediate}
\rho_{\text{th}}=\frac{\bm{\nabla}T_2}{\bm{j}_{q1}}=\left[\frac{Ts^2}{k_{\text{th}}} + 2\kappa\right]^{-1}. 
\end{equation}
When deriving this result we took into account that the thermal current vanishes in the passive layer, $\bm{j}_{q2}=0$, which fixes the hydrodynamic velocity induced by the drag, $\bm{v}_2=(\kappa/Ts)\bm{\nabla}T_2$. This is then used in the force balance condition, $k_\text{th}(\bm{v}_1-\bm{v}_2)=s\bm{\nabla}T_2$, to relate $\bm{v}_1$ and $\bm{\nabla} T_2$. Substituting Eq.~\eqref{eq:kth-result} into~\eqref{eq:rth-intermediate} and observing that in the hydrodynamic regime the second term in the square bracket above may be neglected in comparison with the first, we get a linear relation between the thermal drag resistivity and the drag friction coefficient, 
\begin{equation}\label{eq:kth}
k_{\text{th}}=Ts^2\rho_{\text{th}}.     
\end{equation}
This relation does not rely on the specific form of the drag friction coefficient in Eq.~\eqref{eq:kth-result}, and holds as long as the drag thermal resistivity is small in comparison with the intrinsic thermal resistivity of the layer, $\rho_{\text{th}} \kappa \ll 1$. For the specific form of the friction coefficient in Eq.~\eqref{eq:kth-result} we get
\begin{equation}\label{eq:rth}
\rho_{\text{th}}=\frac{\zeta(3)}{64\pi^2}\left(\frac{\varepsilon}{\sigma s^2d^3}\right).
\end{equation}   
In Sec. \ref{sec:estimates} we provide estimates for the temperature dependence of $k_{\text{th}}$ for both MLG and BLG systems.

\subsection{Near-field heat transfer conductance}

In longer systems the temperature gradients arising in the layers cause appreciable heat transfer between the layers. At small temperature differences the interlayer heat flux is proportional to the temperature difference between the layers, and can be characterized by a thermal conductance per unit area. This near-field thermal conductance can be readily evaluated using the formalism developed above. For that purpose, we consider a situation in which one layer is hotter than the other $T_1>T_2$ and evaluate the energy flux between the layers by computing the work per unit time done by the density fluctuations in the hot layer on the electrons in the cold layer. Using Ehrenfest's theorem~\cite{LL-V5}, $\frac{d}{dt}\langle {\Hat{H}}\rangle = \langle \partial_t \Hat{H} \rangle$,  we can write the heat flux per unit area in the form
\begin{equation}
\mathcal{J}_E= \frac{e}{2} \langle  \delta n_2\partial_t\delta\phi_2 - \delta n_1\partial_t\delta\phi_1\rangle .
\end{equation}
After Fourier transform this expression can be equivalently rewritten as follows 
\begin{equation}
\mathcal{J}_E=\int\frac{d^2qd\omega}{(2\pi)^3} \frac{-i\omega}{2} \left(\frac{2 \pi e^2}{\varepsilon q}\right)e^{-qd}\mathcal{D}_E(\bm{q},\omega),
\end{equation}
where $\mathcal{D}_E$ is defined in analogy with Eq.~\eqref{eq:D} except that it is expressed in the basis of symmetrized densities, $\mathcal{D}_E=\langle\delta n_+\delta n_-\rangle$. To determine the correlation function of fluctuating densities, we use Eq.~\eqref{eq:delta-n} and generalize Eq.~\eqref{eq:j-j} to the situation of layers kept at different temperatures. As a result, we obtain 
\begin{equation}
  \mathcal{D}_E(\bm{q},\omega)=\Delta T\frac{\sigma}{2 e^2}\frac{q^2}{\Gamma^*_+\Gamma_-},
\end{equation}
where $\Delta T = T_1-T_2$ is the temperature difference between the layers.
Upon the frequency integration in $\mathcal{J}_E$, we get 
\begin{equation}
\mathcal{J}_E= - \Delta T \frac{\sigma}{e^2}\int\frac{q^2d^2q}{(2\pi)^2} \,  \left(\frac{ \pi e^2}{\varepsilon q}\right) e^{-qd}\frac{\gamma_+-\gamma_- }{\gamma_+ + \gamma_-}.
\end{equation}
Performing the momentum integration, we obtain the near-field thermal conductance per unit area 
\begin{equation}\label{eq:kappa-th}
\varkappa_\text{th}=-\frac{\mathcal{J}_E}{\Delta T}=\frac{\sigma}{8\varepsilon d^3}. 
\end{equation}

We would like to make several comments on the results presented
in this and the preceding sections: (i) 
The hydrodynamic description of the relevant density fluctuations assumes that their wavelengths, $\sim d$, are longer than the inelastic relaxation length, 
and their frequencies are smaller than the equilibration rate of the electron liquid. Since the electron liquid in graphene is strongly coupled, this assumption is justified when $d$ exceeds the thermal de Broglie wavelength of the electrons.
(ii) For MLG, our results differ from the results obtained using the random-phase approximation (RPA) treatment of density fluctuations \cite{Ying-Kamenev}. We believe that the discrepancy arises because collision-induced damping is neglected in the RPA treatment. For the strongly interacting liquid in graphene this assumption is not expected to hold. 
(iii) Our result for the near-field thermal conductivity in Eq.~\eqref{eq:kappa-th} is expressed in terms of the electrical conductivity on the liquid. Since charge transport is decoupled from the hydrodynamic flow at charge neutrality,  this suggests that the results apply beyond the hydrodynamic regime. Indeed, our derivation only relied on the assumption that the relevant charge-density fluctuations obey Maxwellian relaxation, controlled by the layer  conductivity. In Appendix \ref{sec:appendix} we present an alternative treatment of near-field thermal conductivity, which is in similar spirit to the method developed by E. M. Lifshitz for the evaluation of van der Waals forces between solids \cite{lifshitz1955theory,lifshitz1992theory}, and may be applied to a double-layer of conductors of arbitrary thickness.   

\begin{figure}[t!]
\includegraphics[width=\linewidth]{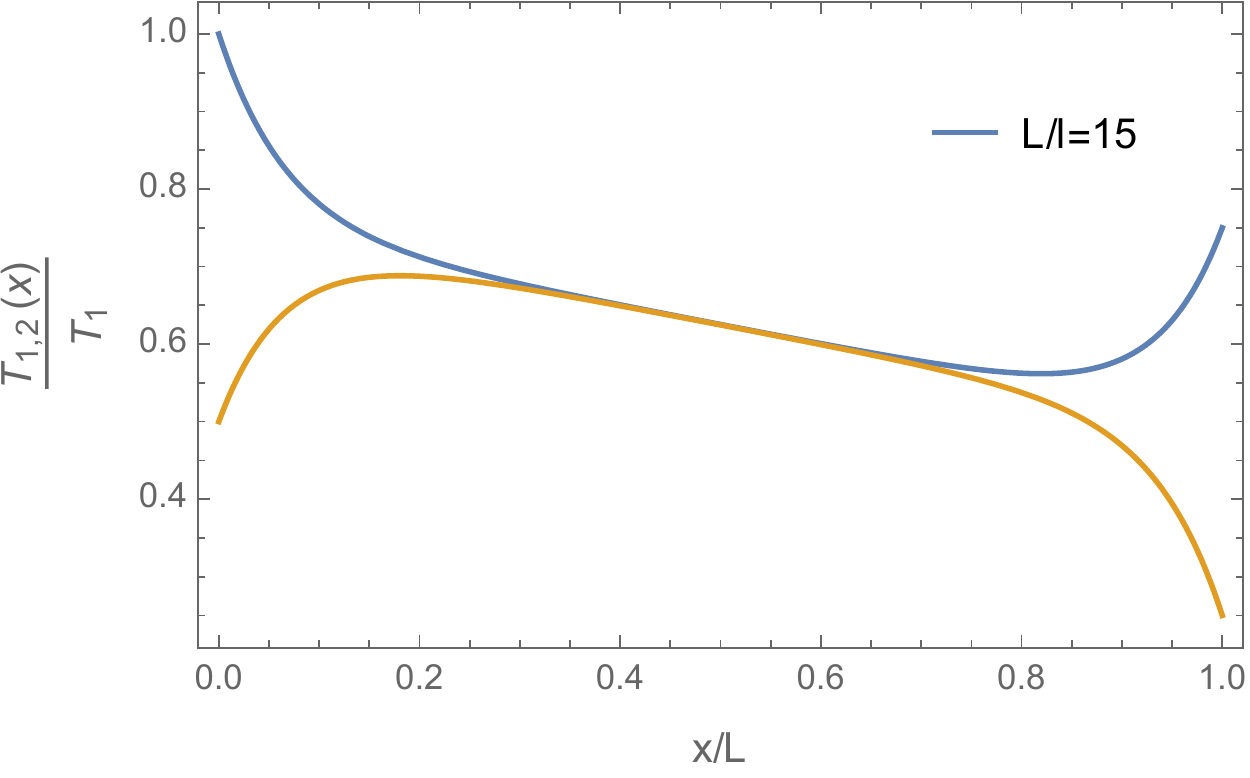}
\caption{The spatial profile of the temperature distributions in the EDL under the regime of the thermal drag effect. The plot is generated from Eq. \eqref{eq:T12}
for the thermal healing length satisfying the condition $L/l=15$ and the following choice of parameters: $\alpha=3/4$, $\beta=1/2$, and $\tau=1/2$.}
\label{fig:T12}
\end{figure}

\subsection{Thermal transport in long systems: Thermal locking and perfect drag}  

As discussed above, in sufficiently long systems both thermal drag and near-field interlayer thermal conductivity significantly affect thermal transport in double-layers. In this section we develop a macroscopic description of the thermal transport in EDL  by accounting for the effects of both thermal drag and near-field heat transfer between the layers. 

Thus far we have considered perfectly clean systems,  whereas realistic samples are disordered. To account for the presence of disorder we assume that each layer is subject to the disorder potential with a long (in comparison to the equilibration length of the electron liquid) correlation radius $\xi$. There are multiple pieces of experimental evidence in support of this model assumption. Indeed, scanning probe microscopy on hBN encapsulated graphene reveals electron-hole charge puddles with the typical correlation radius in the range of $\xi\sim 100$ nm and local strength of $U \sim 5$ meV; see Refs. \cite{Crommie,LeRoy}. One should note that, in this model, the disorder-induced momentum relaxation in the flow may not be described in terms of individual collisions of electrons with impurities since microscopic length scale of momentum-conserving electron collisions is much shorter than $\xi$. Instead the momentum relaxation in this regime must be described by the hydrodynamic approach~\cite{AKS}. For the electron liquids in graphene this was done in Refs. \cite{Lucas-PRB16,Li-PRB20}. In this approach, a description of the transport in the system at spatial scales exceeding $\xi$ is obtained by averaging the hydrodynamic equations in an external potential over disorder realizations. As a result, the force balance equation Eq.~\eqref{eq:dt-p} acquires a friction term \cite{Li-PRB20}
\begin{equation}\label{eq:friction}
\bm{F}=-k\bm{v},\quad k=\frac{\left\langle(s\delta n-n\delta s)^2\right\rangle}{2\left(\frac{n^2\kappa}{T}-\frac{2ns\gamma}{T}+\frac{s^2\sigma}{e^2}\right)}.
\end{equation}
In doped systems the friction coefficient $k$ in Eq. \eqref{eq:friction}
is determined by the entire matrix of thermoelectric coefficients [Eq. \eqref{eq:Upsilon}] as well as both particle density $\delta n(\bm{r})$ and entropy density $\delta s(\bm{r})$ fluctuations. In the regime near charge neutrality, $n\to0$, its functional form simplifies to $k=\frac{e^2}{2\sigma}\langle\delta n^2\rangle$. For simplicity, we assume that disorder does not induce interlayer correlations. This assumption is certainly justified at large interlayer separations.

At this point, consider the four-terminal setup of a thermally biased EDL. For the system with the length $L$ we take 
for the lower layer $T_1(x=0)=T_1$ and $T_1(x=L)=\alpha T_1$, and similarly for the top layer $T_2(x=0)=T_2$ and $T_1(x=L)=\beta T_2$, where the parameters $\alpha$ and $\beta$ are  close to unity. We are interested in describing the resulting spatially inhomogeneous flow of heat in the $x$ direction. Since the transfer of energy from the electrons to the phonons is suppressed by the small ratio of the speed of sound to the electron velocity, we neglect the extrinsic energy losses to phonons. This energy relaxation channel can be described by introducing the corresponding electron-lattice thermal conductivity into our equations. 

The entropy current in each layer consists of two terms generated by hydrodynamic velocity and local temperature gradient    
\begin{equation}\label{eq:j-s}
\bm{j}_{si}=s\bm{v}_i-\frac{\kappa}{T}\bm{\nabla}T_i,\quad i=1,2.
\end{equation}
The entropy conservation is formulated in terms of the continuity equation that states the loss of the entropy current in one layer is governed by the outflow into the other layer via the NFHT effect     
\begin{subequations}
\label{eq:heat_current_continuity}
\begin{align}
\partial_x j_{s1}+\frac{\varkappa_{\text{th}}}{T}(T_1-T_2)=0, \\ 
\partial_x j_{s2}-\frac{\varkappa_{\text{th}}}{T}(T_1-T_2)=0.
\end{align}
\end{subequations}
In a steady state, the force balance condition in each of the layers is determined by the thermal and frictional forces  
\begin{subequations}
\label{eq:force_balance}
\begin{align}
-k v_1-k_{\text{th}}(v_1-v_2)-s\partial_xT_1=0, \\ 
-k v_2-k_{\text{th}}(v_2-v_1)-s\partial_xT_2=0,
\end{align}
\end{subequations}
which include both the disorder-induced friction Eq. \eqref{eq:friction} and the thermal drag friction Eq. \eqref{eq:kth}. 

To solve this system of linear equations it is advantageous to rewrite them in the symmetrized representation.  For instance, for temperatures $T_\pm(x)=T_1(x)\pm T_2(x)$ we find 
\begin{subequations}
\label{eq:T_eqs}
\begin{align}
\left(\frac{Ts^2}{k}+\kappa\right)\partial^2_xT_+=0, \\ 
\left(\frac{Ts^2}{k+2k_{\text{th}}}+\kappa\right)\partial^2_xT_- - 2\varkappa_{\text{th}}T_-=0. 
\end{align}
\end{subequations}
We see that symmetric and antisymmetric parts of the temperature distribution decouple. The former one describes the profile in the bulk of the flow, whereas the latter one captures thermal redistribution near the sample boundaries. The interplay of the gradient term and the local term in the equation for $T_-$ introduces the natural healing length of interlayer thermalization, 
\begin{equation}\label{eq:l}
l^{-2}=\frac{2\varkappa_{\text{th}}}{\kappa+Ts^2/(k+2k_{\text{th}})}.
\end{equation}
Assuming $k \lesssim k_{\mathrm{th}}$, and using Eqs.~\eqref{eq:kth-result} and \eqref{eq:kappa-th} we get the following estimate:
\begin{equation}\label{eq:l_estimate}
\frac{d^2}{l^2} \sim \frac{ 1 }{s^2 d^4} \ll 1,
\end{equation}
which justifies our local approximation. As shown in the following section this condition can be easily met in graphene devices.

The solution of Eqs.~\eqref{eq:T_eqs} satisfying the boundary conditions is given by
\begin{equation}\label{eq:T12}
\frac{T_{1,2}(x)}{T_1}=\frac{1}{2}[f_a(x)\pm f_b(x)\pm f_c(x)].
\end{equation} 
Here the upper (lower) sign is for $T_1(x)/T_2(x)$ and we introduced three dimensionless functions 
\begin{subequations}
\begin{align}
&f_a(x)=(1+\tau)\left(1-\frac{x}{L}\right)+(\alpha+\beta\tau)\frac{x}{L}, \\ 
&f_b(x)=(\alpha-\beta\tau)\frac{\sinh(x/l)}{\sinh(L/l)},\\
&f_c(x)=(1-\tau)\frac{\sinh((L-x)/l)}{\sinh(L/l)},
\end{align}
\end{subequations}
where $\tau=T_2/T_1$. To illustrate the effect of interlayer heat transfer and competing thermal drag processes we plot temperature profiles in Fig. \ref{fig:T12} for a particular choice of parameters. In the bulk of the flow, far away from the system boundaries, where $l\ll x\ll L$, we can clearly identify thermally locked state,  where the local temperatures in the two layers become identical. This locking is caused by interlayer thermal conductivity, which, in combination with force balance, leads to exponential decay of interlayer temperature difference on the scale of the healing length $l$ defined by Eq.~\eqref{eq:l}.  From the linear combination of the force balance equations~\eqref{eq:force_balance}, we determine the spatial gradient of the locked temperatures in the bulk of the device, $-k v_+-s\partial_x T_+=0$. Thus, the temperature gradient in the bulk of the flow  determines hydrodynamic velocities, $\bm{v}_1\approx\bm{v}_2$, with exponential accuracy in the thermal healing length,  $\bm{v}_i\approx -(s/k)\bm{\nabla}T_i$. Therefore,  the entropy currents defined by Eq.~\eqref{eq:j-s} in the bulk of the flow, $l\ll x\ll L$, can be expressed as follows   
\begin{equation}\label{eq:perfect-drag}
\bm{j}_{si}=-\left(\frac{Ts^2}{k}+\kappa\right)\frac{\bm{\nabla} T_i}{T}=-\mathcal{K}\frac{\bm{\nabla} T_i}{T}.
\end{equation} 
The coefficient $\mathcal{K}$ can be considered as an effective thermal conductivity in the regime of the perfect drag effect. This equation coincides with the expression for the thermal conductivity of the electron liquid in the presence of disorder. This reflects the fact that in the regime of thermal locking the double-layer acts as a single electron liquid with identical hydrodynamic velocities in the two layers. 

\subsection{Estimates}\label{sec:estimates}

In this section we provide estimates for the obtained kinetic coefficients in graphene EDL devices. 

For MLG devices the intrinsic conductivity is known to be of the order of conductance quantum $\sim e^2/2\pi$ (in units of $\hbar=1$) modulo logarithmic renormalizations in the weak-coupling theory \cite{Fritz,Kashuba}:
\begin{equation}
\sigma=\frac{e^2}{2\pi\alpha^2_T},\quad \alpha_T=\frac{\alpha_g}{1+(\alpha_g/4)\ln(\Lambda/T)}, 
\end{equation}
where $\alpha_g=e^2/(\varepsilon v)$ is the temperature-independent dimensionless interaction constant, which determines the bare strength of the electron-electron interactions, and $\Lambda$ is the cutoff in the scheme of the renormalization group. The entropy density of MLG at charge neutrality can be estimated as $s\sim (T/v)^2$. Therefore, neglecting all logarithmic factors, we find from Eq.~\eqref{eq:kappa-th} that the NFHT conductance $\varkappa_{\text{th}}$ is only weakly (logarithmically) temperature dependent. For the thermal drag resistance in Eq.~\eqref{eq:rth} we get $\rho_{\text{th}}\propto 1/T^4$. For the healing length we deduce from Eq. \eqref{eq:l_estimate} that $d/l \sim (\lambda_T/d)^2\ll1$. 

Furthermmore, it is useful to compare the resulting thermal drag friction coefficient $k_\text{th}$ to the disorder-induced friction coefficient $k$. For that we need to express the density variance $\langle\delta n^2\rangle$ in terms of the properties of the disorder potential $U$. This can be done in the linear screening approximation, where the equilibrium density modulation is related to the external potential as $\delta n(q)=-\nu qU(q)/(q+r^{-1}_{\text{TF}})$, with $r_{\text{TF}}=1/(2\pi e^2\nu)$ being the Thomas-Fermi screening radius and $\nu\sim T/v^2$ is the thermodynamic single-particle density of states. This can be further simplified by noticing that in the hydrodynamic regime, the correlation radius of disorder $\xi$ exceeds the Thomas-Fermi screening radius $r_{\text{TF}}$. Therefore, $k\sim (e^2/\sigma)\langle U^2\rangle/(\xi^2e^4)$, where we additionally assumed that the spectral density of disorder potential does not have strong divergence at $q\to0$ (e.g., encapsulation-induced disorder). Putting everything together, we deduce 
\begin{equation}
\frac{k_{\text{th}}}{k}\sim\left(\frac{e^2}{\varepsilon v}\right)\left(\frac{\xi}{d}\right)^2\frac{TT_d}{\langle U^2\rangle},\quad T_d=v/d.
\end{equation}     
Taking $\xi\sim 100$ nm and $d\sim200$ nm we estimate $T_d\sim 50$K with the typical velocity $v\sim10^6$ m/s. As a result, for the typical potential variations in the scale of $U\sim5$ meV, we have $k_\text{th}\sim k$ for the range of temperatures $T\sim 50-100$ K wherein hydrodynamic effects are most pronounced. 

Similar estimates can be obtained for BLG devices.
For instance, the entropy density of BLG has different temperature dependence, $s\sim m^* T$, where $m^*$ is the effective mass of the band structure. As a consequence, one expects different results for the resistance, $\rho_{\text{th}}\propto 1/T^2$. On the other hand, the NFHT conductance remains parametrically the same, since the intrinsic conductivity of BLG $\sigma(n=0,T)$ is expected to display rather weak temperature dependence \cite{Morpurgo}. 

\section{Summary and perspective}\label{sec:summary}

In this work we studied thermal transport in electronic double-layers in the regime of global charge neutrality, where transport of charge decouples from that of heat. In these systems thermal transport is dominated by thermally induced density fluctuations that are coupled by the interlayer Coulomb forces.  This produces two effects: (i) thermal drag, described by the corresponding resistivity in Eq.~\eqref{eq:rth}; and (ii) interlayer near-field heat transfer, described by the conductivity in Eq.~\eqref{eq:kappa-th}. One of the key conclusions that follow from our analysis is the fact that, in systems without Galilean invariance, the drag is much stronger than in Galilean-invariant systems, and has a much slower falloff with interlayer distance. Another key finding concerns the regime of perfect drag. The spatial dependence of temperatures and the hydrodynamic velocity in the two layers is determined by the interplay between drag and NFHT processes. For systems that are longer than healing length $l$ for interlayer thermalization in Eq.~\eqref{eq:l}, the bulk flow corresponds to the thermally locked state which is characterized by the same hydrodynamic velocity in each layer. In this regime EDL acts as a single electron liquid with an effective thermal conductivity defined by Eq.~\eqref{eq:perfect-drag}.   

Earlier studies revealed that the decoupling of thermal and electrical transport leads to anomalous thermoelectric responses near charge neutrality. The latter is most notably manifested by the observed gross violation of the Wiedemann-Franz law as captured by the Lorenz ratio \cite{Crossno} and Mott relation for the Seebeck coefficient \cite{Ghahari}. Therefore, transport in EDLs open additional avenues for the exploration of related thermal transport phenomena. For instance, although we focused our consideration on electron double-layers, the general theory of hydrodynamic fluctuations in electron systems without Galilean invariance developed above, is applicable to other device geometries. In this regard, we would like to mention that various drag anomalies were observed in other hybrid circuits that include carbon nanotube-MLG and InAs quantum wire-MLG double-layers \cite{Das,Kim}, where our theory may find useful applications.

\section*{Acknowledgments}

We acknowledge Aaron Hui and Brian Skinner for discussions on the topic of hydrodynamic fluctuations in electron liquids. We thank Philippe Ben-Abdallah for reading and commenting on the paper. This work was financially supported by the National Science Foundation Grants No. DMR-2203411 (A. L.) and DMR-1653661 (S.L.), and MRSEC Grant No. DMR-1719797 (A.V.A.). This project was finalized during the conference “Quantum and Thermal Electrodynamic Fluctuations in the Presence of Matter: Progress and Challenges” at the Kavli Institute for Theoretical Physics, where this research was supported in part by the National Science Foundation under Grant No. NSF PHY-1748958.

\appendix 

\section{Langevin approach}\label{sec:appendix}

The derivation of the near-field thermal conductivity presented in the main text relied only on the Maxwellian relaxation mechanism of relevant density fluctuations. In this appendix we present an alternative consideration of near-field interlayer thermal conductance, which applies to conductors of arbitrary thickness. For conductors of finite thickness the charge fluctuations can spread not only along the plane but also in the perpendicular direction. 
We then apply this approach to reproduce the known results for the near-field conductance between two semi-infinite metals, which has a drastically different dependence on the interlayer distance, temperature, and the conductivity of the metals. 

Fluctuations of electric current $\bm{j}$ and electric field $\bm{E}$ in conductors can be treated within the scheme of Langevin approach \cite{Kogan},
\begin{subequations}
\begin{align}
\bm{j}_\omega=\sigma\bm{E}_\omega+\delta\bm{j}_\omega,\quad \bm{\nabla}\cdot\bm{E}_\omega=\frac{4\pi}{i\omega}\bm{\nabla}\cdot\bm{j}_
\omega, \\ 
\langle\delta\bm{j}_\omega(\bm{r})\delta\bm{j}_{-\omega(}\bm{r}')\rangle=\frac{\omega^2\ImP\varepsilon_\omega}{2\pi}\coth\frac{\omega}{2T}\delta(\bm{r}-\bm{r}').
\end{align}
\end{subequations}
Here we assumed absence of spatial dispersion of the dielectric constant, $\varepsilon_\omega=1+4\pi i\sigma/\omega$. We also consider only the fluctuations of the longitudinal electric field. Assuming that the gap size $d$ is much smaller than the thermal wavelength, $\lambda_T=2\pi c/T$, the contribution of transverse fluctuations should be smaller in $d/\lambda_T\ll1$.   

The longitudinal current fluctuations can be expressed in terms of the fluctuations of polarization, $\delta\bm{P}_\omega(\bm{r})=i\delta\bm{j}_\omega(\bm{r})/\omega$. We will therefore use intrinsic fluctuations of polarization as the Langevin source. Denoting the fluctuations of the scalar potential by $\delta\phi$ we obtain an inhomogeneous partial differential equation obeyed by them, 
\begin{equation}\label{eq:phi-P}
\nabla^2\delta\phi_\omega+\frac{4\pi i}{\omega}\bm{\nabla}\cdot(\sigma\bm{\nabla}\delta\phi_\omega)=-4\pi\bm{\nabla}\cdot\delta\bm{P}_\omega(\bm{r}).    
\end{equation}
This framework parallels with the earlier discussion presented in Sec. \ref{sec:hydro} for electron liquids in hydrodynamic regime. Instead, here we tailor the formalism to describe thermal electrodynamic fluctuations in the presence of matter for a generic conductor. 

\subsection{Nanogap geometry}

To find the contribution of electric-field fluctuations to the thermal conductivity of a nanogap device we consider two semi-infinite conductors filling the space at $|x|>d$ separated by a vacuum gap, $|x|<d$. Owing to the translation invariance of the problem in $yz$ plane we use Fourier representation for the $y$ and $z$ variables. In this representation the resolvent corresponding to Eq. \eqref{eq:phi-P}, the Green's function, obeys the equation 
\begin{equation}\label{eq:G}
\left[\partial^2_x+\frac{4\pi i}{\omega}\partial_x\sigma(x)\partial_x-q^2\varepsilon_\omega\right]G_\omega(x,x')=\delta(x-x')    
\end{equation}
whose solution can be expressed as 
\begin{equation}
G_\omega(x,x')=\frac{1}{W(x')}\left\{\begin{array}{cc}u(x)v(x') & x<x' \\ v(x)u(x') & x>x'\end{array}\right.    
\end{equation}
where $u(x)$ and $v(x)$ are two independent solutions of Eq. \eqref{eq:G} with the vanishing right-hand side that vanish respectively at $x\to-\infty$ and $x\to+\infty$, and $W$ is their Wronskian, 
\begin{equation}
W(x)=\left|\begin{array}{cc}u(x) & v(x) \\ u'(x) & v'(x)\end{array}\right|    
\end{equation}
In what follows we consider a homogeneous junction with identical conductors with the conductivity $\sigma$ on both sides of the junction, namely $\sigma(x)=\sigma\theta(x^2-d^2)$. In this case the problem is symmetric with respect to reflection in the $x=0$ plane and the two solutions of the homogeneous equation are related by $v(x)=u(-x)$. We also recall that, for the bulk metal case,
the three-dimensional conductivity $\sigma$ has the dimensionality of the inverse time, so that $1/(4\pi\sigma)$ (in CGS units) has a meaning of the $RC$ time needed to dissolve a charge-density perturbation.

In general, the Green's function can be defined for an arbitrary layer thickness. For thin layers, the second term in the square bracket in (A3) may be dropped. This corresponds to neglecting polarization of the planes perpendicular to the layers. In this case the approach reproduces our results from the main text. The opposite limit of semi-infinite conductors is considered below. The method enables consideration of the entire crossover between these two limiting cases. 

\subsection{Green's function}

Inside the left conductor, $x<-d$, the first solution is $u(x)=e^{qx}$. Inside the gap it can be written as the linear superposition of two exponentials 
\begin{equation}
u(x)=\mathcal{T}_\omega e^{qx}+(1-\mathcal{T}_\omega)e^{-2qd-qx}, \quad |x|<d.    
\end{equation}
The value of the transmission amplitude $\mathcal{T}_\omega$ is obtained from the boundary condition,
\begin{equation}
u'(-d+\eta)-u'(-d-\eta)=\frac{4\pi i\sigma}{\omega}u'(-d-\eta),
\end{equation}
where $\eta\to0$ is a positive infinitesimal. This gives 
\begin{equation}
\mathcal{T}_\omega=1+\frac{2\pi i\sigma}{\omega}.    
\end{equation}
Inside the right conductor we write the first solution in the form 
\begin{equation}
u(x)=\mathcal{A}_\omega e^{qx}+\mathcal{B}_\omega e^{-qx}, \quad x \geq d.
\end{equation}
The values of the constants $\mathcal{A}_\omega$ and $\mathcal{B}_\omega$ are found from the continuity of $u(x)$ at $x=d$ and from the condition on the discontinuity of its derivative, which follows from Eq. \eqref{eq:G},
\begin{equation}
u'(d+\eta)-u'(d-\eta)=-\frac{4\pi i\sigma}{\omega}u'(d+\eta).
\end{equation}
This gives 
\begin{subequations}\label{eq:A-B}
\begin{align}
&\mathcal{A}_\omega=\frac{1}{\varepsilon_\omega}\left[\mathcal{T}^2_\omega-(1-\mathcal{T}_\omega)^2e^{-4qd}\right], \\ 
&\mathcal{B}_\omega=-\frac{1}{\varepsilon_\omega}\mathcal{T}_\omega(1-\mathcal{T}_\omega)\sinh(qd).
\end{align}    
\end{subequations}
The coordinate dependence of the Wronskian can be found to have a form 
\begin{equation}
W(x)=-2q\mathcal{A}_\omega
\left\{\begin{array}{cc}1&|x|\geq d \\ \varepsilon_\omega & |x|<d\end{array}\right.
\end{equation}
Below we will need the Green's function in the region $x<-d$ and $x>d$, where it is given by 
\begin{equation}
G_\omega(x,x')=-\frac{1}{2q\mathcal{A}_\omega}e^{q(x-x')} .   
\end{equation}

\subsection{Heat fluxes and NFHT conductance}

The heat flux from the right conductor to the left one can be found as follows. The fluctuations of the electronic field in the left conductor induced by the Langevin sources in the right conductor are described by, 
\begin{equation}\label{eq:delta-phi}
\delta\phi_\omega=\frac{2\pi}{q\mathcal{A}_\omega}\mathrm{e}^{qx}\int^{\infty}_{d}dx'\left[iq\delta\mathcal{P}^\perp_\omega(x')+\partial_x\delta\mathcal{P}^x_{\omega}(x')\right]e^{-qx'}.
\end{equation}
The hear flux from the right conductor to the left equals the Joule -heat losses induced by the above fluctuations inside the left conductor. For a nanogap device with a surface area $S$ they are given by 
\begin{equation}
\dot{Q}^R_{R\to L}=2S\sigma\int\frac{d\omega d^2q}{(2\pi)^3}\int^{-d}_{-\infty}dx q^2\langle\delta\phi_\omega(x)\delta\phi_{-\omega}(x)\rangle.
\end{equation}
Taking the variance of Langevin sources corresponds to thermal equilibrium at temperature $T_R$,
\begin{equation}
\langle\delta\mathcal{\bm{P}}_\omega(\bm{r})
\delta\mathcal{\bm{P}}_{-\omega}(\bm{r}') \rangle=\frac{2\sigma}{\omega}\coth\frac{\omega}{2T_R}\theta(x-d)\delta(\bm{r}-\bm{r}'),
\end{equation}
we obtain 
\begin{align}
\langle[iq\delta\mathcal{P}^\perp_\omega(x)+\partial_x\delta\mathcal{P}^x_\omega(x)][-iq\delta\mathcal{P}^\perp_{-\omega}(x')+\partial_{x'}\delta\mathcal{P}^x_{-\omega}(x')]\rangle\nonumber \\ 
=\frac{2\sigma}{\omega}\coth\frac{\omega}{2T_R}[q^2+\partial_x\partial_{x'}]\theta(x-d)\delta(x-x')
\end{align}
As the next step, we use Eq. \eqref{eq:delta-phi} and obtain 
\begin{align}
&\dot{Q}_{R\to L}=2S\sigma^2\int\frac{d\omega}{2\pi}\frac{\coth\frac{\omega}{2T_R}}{\omega}\int\frac{d^2q}{q}\frac{e^{-2qd}}{|\mathcal{A}_\omega|^2}
\nonumber\\ 
&\iint^{\infty}_{d} dx dx' e^{-q(x+x')}[q^2+\partial_x\partial_{x'}]\theta(x'-d-\eta)\delta(x-x')
\end{align}
The presence and the sign of the infinitesimal $\eta$ in the argument of the theta function is forced by the choice if the value of the Wronskian corresponding to $x'>d$. Physically this means that the surface charge arising from Langevin sources should be viewed as residing inside the conductor. Performing the integration over coordinates $x$ and $x'$ in the last expression we find
\begin{equation}
\dot{Q}_{R\to L}=3S\sigma^2\int\frac{d\omega}{2\pi}\frac{\coth\frac{\omega}{2T_R}}{\omega}
\int^{\infty}_{0}qdq\frac{e^{-4qd}}{|\mathcal{A}_\omega|^2}.
\end{equation}
In this expression, quantum noise needs to be disregarded as it cannot produce dissipation. It is temperature independent and will cancel with the opposite heat flux $\dot{Q}_{L\to R}$ leading to the net energy current density $\mathcal{J}_E=[\dot{Q}_{R\to L}-\dot{Q}_{L\to R}]/S$. The important frequencies are of the order of the temperature $\omega \sim T$. Using Eq. \eqref{eq:A-B}   
and assuming $T\ll\sigma$ we can estimate with logarithmic accuracy 
\begin{equation}
\varkappa_{\text{th}}=\lim_{T_{1,2}\to T}\frac{\mathcal{J}_E(T_1,T_2,d)}{T_1-T_2}\simeq\frac{\pi^3}{160} \frac{T}{d^2}\left(\frac{T}{\sigma}\right)^2\ln\frac{\sigma}{T}.  
\end{equation}
This analysis reproduces known results form the earlier studies \cite{Polder-VanHove,Basko-NFHT} obtained in the same limit. In particular, we see that in three-dimensional devices both temperature and interlayer separation dependencies are different as compared with two-dimensional case considered in the main part of this work.  

\bibliography{biblio}

\end{document}